\documentclass[aps,prb,twocolumn,amsmath,amssymb,showpacs]{revtex4-1}
\usepackage{graphicx}
\usepackage{dcolumn}
\usepackage{bm}

\begin{document}

\title{Non-Abelian gauge fields in the gradient expansion:
\\generalized Boltzmann and Eilenberger equations}

\author{C. Gorini$^a$, P. Schwab$^{a,b}$, R. Raimondi$^c$, and A. L.  Shelankov$^{d,e}$}
\affiliation{$^a$Institut f\"ur Physik, Universit\"at Augsburg, 86135 Augsburg, Germany\\
$^b$Institut f\"ur Mathematische Physik, Technische Universit\"at, Braunschweig, Germany\\
$^c$CNISM and Dipartimento  di Fisica "E. Amaldi", Universit\`a  Roma Tre, 00146 Roma, Italy\\
$^d$Department of Physics, Ume\aa~ University, Ume\aa, Sweden\\
$^e$Ioffe Physicotechnical Institute of the RAS, St. Petersburg, Russia
}

\date{\today}

\begin{abstract}
We present a microscopic derivation of the generalized Boltzmann and Eilenberger
equations in the presence of non-Abelian gauges, for the case of a non-relativistic disordered Fermi gas.  
A unified and symmetric treatment of the charge $[U(1)]$ and spin $[SU(2)]$ degrees of freedom
is achieved.  Within this framework, just as the $U(1)$ Lorentz force generates the Hall effect, 
so does its $SU(2)$ counterpart give rise to the spin Hall effect.
Considering elastic and spin-independent disorder
we obtain  diffusion equations for charge and spin densities and show
how the interplay between an in-plane magnetic field and a time dependent Rashba
term generates in-plane charge currents.
\end{abstract}

\maketitle

\section{Introduction}
Spin-charge coupled dynamics in two-dimensional electron (hole) gases
has been the focus of much theoretical and experimental work over the last
two decades \cite{zutic2004,awschalom2007}.  
Its rich physics belongs to the field of spintronics
and shows much potential for applications.
Thanks to spin-orbit coupling all-electrical control of
the spin degrees of freedom of carriers, as well as magnetic
control of the charge one, is in principle possible \cite{inoue2009}.
Particularly interesting from this point of view are phenomena 
like the spin Hall effect and the anomalous Hall effect.   
For a review of both, see Ref.~[\onlinecite{engel2007}] and 
Refs.~[\onlinecite{sinitsyn2008b,nagaosa2009}], respectively.
In general terms the theoretical problem at hand is that
of describing spin-charge coupled transport in a disordered system.
In the semiclassical regime, defined by the condition $\lambda_F\ll l$,
a Boltzmann-like treatment is sensible and expected to provide
physical transparency.
Here $\lambda_F$ is the Fermi wavelength and $l$ a typical length scale
characterizing the system -- say, the mean free path or that defining
spatial inhomogeneities due to an applied field.
The Boltzmann equation is a versatile and powerful tool for the description
of transport phenomena\cite{ziman},
and various generalizations to the case in which
spin-orbit coupling appears have been proposed \cite{khaetskii2006, trushin2007, kailasvuori2009}. 
More general Boltzmann-like equations have also
been obtained \cite{mishchenko2004, shytov2006, raimondi2006}.
In both cases though, much of the physical transparency is lost due
to a complicated structure of the velocity operator and of the collision integral.
A semiclassical approach based on wave packet equations \cite{sundaram1999, sinitsyn2007}
can partially circumvent these complications, though it is limited to
the regime $\Delta_{so}\tau/\hbar\gg1$, with $\Delta_{so}$ the spin-orbit energy
and $\tau$ the quasiparticle lifetime.   
On the other hand it was pointed out
in different works \cite{mathur1992, frohlich1993, jin2006, bernevig2006, hatano2007, tokatly2008, tokatly2010} 
that Hamiltonians with a linear-in-momentum spin-orbit coupling term can be
treated in a unified way by introducing $SU(2)$ gauge potentials
in the model.
Taking as an example the Rashba Hamiltonian
\begin{equation}
H_{R} =
\frac{p^2}{2m} + \alpha\left(p_y\sigma^x-p_x\sigma^y\right),
\end{equation}
where $\alpha$ is the spin-orbit coupling constant, one can identically transform it to
\begin{equation}
\label{rashba0}
H_{R} =
\frac{\left[{\bf p}+\gamma\boldsymbol{\mathcal A}_{R}^a\sigma^a/2\right]^2}{2m} + const.
\end{equation}
Here summation over $a=x,y,z$ is implied, $\gamma$ is the $SU(2)$ coupling constant,
and the components of the $SU(2)$ vector potential 
are $\gamma({\mathcal A}_{R})^x_y=-\gamma({\mathcal A}_{R})^y_x=2m\alpha,
({\mathcal A}_R)^x_x=({\mathcal A}_R)^y_y=0$.  
From this point of view a different Hamiltonian, 
say the Dresselhaus one, simply corresponds to a different choice of the vector potential.
An additional advantage of this approach is that it ensures the proper
definition of physical quantities like spin currents and
polarizations\cite{tokatly2008, duckheim2009}.
More generally, the use of the non-Abelian language shows flexibility and
potential, and has already proven useful in different contexts.
For example, in Ref.~[\onlinecite{bernevig2006}] it was used to predict the existence
of a ``persistent spin helix'' in systems with equal strength Rashba and Dresselhaus couplings.  
Such a helix was later observed \cite{koralek2009} 
and soon after exploited \cite{wunderlich2009}.
The authors of Ref.~[\onlinecite{hatano2007}] on the other hand employed it in their proposal
of a perfect spin-filter based on mesoscopic interference circuits.
Finally, since non-Abelian potentials can also 
be created optically \cite{osterloh2005, gunter2009, liu2009, merkl2010},
the range of applications of the approach goes beyond systems
described by Hamiltonians like (\ref{rashba0}).
Indeed, even in solid state systems higher-dimensional models
such as the one considered in Ref.~[\onlinecite{lee2009}] would fit into the picture
\footnote{In this case the Hamiltonian can be written introducing
$SU(4)$ -- rather than $SU(2)$ -- gauge fields.}. 

Our goal in the present paper is therefore to put the non-Abelian
approach in the semiclassical regime on firm ground,
in order to obtain kinetic equations with a clear physical structure and as broad
a field of application as possible.
More precisely, we derive an $SU(2)\times
U(1)$ covariant Boltzmann equation 
in the framework of the Keldysh \cite{Kel64} microscopic formalism.
In the covariant approach a completely symmetric treatment of the
charge and spin degrees of freedom is achieved 
\cite{[{A similar approach in a somewhat different context was pursued in }] shindou2008}.
Also, we discuss the more general Eilenberger 
equation \cite{rammer1986, prange1964, schwab2003} derived with the
help of the so called
$\xi$-integrated Green's function technique. The latter allows one to
justify the Boltzmann equation in the case when the momentum is not a
good quantum number due to impurity or other scattering, and the
notion of particles with a given momentum is ill defined.
The results obtained hold in the metallic regime 
$\epsilon_F\gg \hbar/\tau$, with $\epsilon_F$ the Fermi energy and $\hbar/\tau$
the level broadening due to disorder, and as long the spin splitting due to the 
$[SU(2)]$ gauge fields is small compared to the Fermi energy, $\Delta_{so}\ll\epsilon_F$, 
but for arbitrary values of $\Delta_{so}\tau /\hbar $.
We emphasize that, within this approach, not only the applied
electric and magnetic fields, but also the internal spin-orbit induced
ones can be position and time dependent. 

Our guideline for the present work is the familiar $U(1)$ gauge invariant Boltzmann equation.  
This reads \cite{ziman}
\begin{equation}
\label{boltzmann0}
\left(
\partial_T + \frac{{\bf p}^*}{m}\cdot\nabla_{\bf R}+{\bf F}\cdot\nabla_{{\bf p}^*}
\right)f(T,{\bf R},{\bf p}^*) = I[f],
\end{equation}
where the electron distribution function $f$ at time $T$ and 
position $\bm{R}$ is a function of the gauge invariant {\em kinematic} momentum 
 ${\bf p}^*={\bf p}+e{\bf A}(T,\bm{R})$  (rather than the canonical
momentum $\bm{p}$),
and the Lorentz force ${\bf F}=-e[{\bf E}+({\bf p}^*/m)\wedge{\bf B}]$ appears.
The right hand side of Eq.~(\ref{boltzmann0}) contains the collision integral.

The paper is organized as follows.
In Section~\ref{secgradient}
we start by recalling the quantum derivation of the Boltzmann equation,
which allows us to introduce the general formalism in a purposeful way.
In Sections~\ref{secintegration} the generalized Boltzmann and Eilenberger equations
are obtained.  In Section~\ref{secdiff} the diffusive regime is discussed
and spin-charge coupled diffusion equations are derived.
Finally, Section~\ref{secexamples} shows two example calculations.
The first involves a study of the Bloch equations in the static
limit in the presence of in-plane electric and magnetic fields,
whereas the second is concerned with a novel effect, in which
an in-plane charge current is generated by the interplay
of an in-plane magnetic field and a time dependent Rashba term. 

We use a system of units where the Planck constant $\hbar =1$ and $e=|e|$.


\section{The gradient expansion}
\label{secgradient}

The original quantum-mechanical derivation of the classical
Boltzmann equation by Keldysh \cite{Kel64} has been exploited and extended by
many authors, in particular, by  Langreth \cite{langreth1966} and Altshuler \cite{altshuler1978}.
Since it is very instructive, we outline the procedure following Ref.~[\onlinecite{rammer1986}] and
consider for simplicity's sake the case of free electrons in a perfect lattice.
Our aim will be to generalize it to the non-Abelian case
and to later introduce disorder.
The main character is the Green function in Keldysh space $\check{G}$
\begin{equation}
\check{G} = 
\left(
\begin{array}{cc}
G^R & G^K \\
0 & G^A
\end{array}
\right).
\end{equation}
$G^{R,A}$ are the standard retarded and advanced Green  functions, whereas
$G^K$ is the Keldysh Green  function which carries the statistical information about the
occupation of the energy spectrum.
One starts from the left-right subtracted Dyson (quantum kinetic) equation
\begin{equation}
\label{kinetic0}
-i\left[G_0^{-1}(1,1')\stackrel{\otimes}{,}\check{G}(1',2)\right] = 0,
\end{equation}
where $1, 1', 2$ are generalized coordinates containing space and
time coordinates as well as spin and Keldysh space (and possibly additional) indices. 
The square brackets denote the commutator, the symbol ``$\otimes$''
indicates convolution/matrix multiplication over the internal variables/indices and
\begin{equation}
G_0^{-1}(1,1') = \Big(i\partial_{t_1}-\frac{\left[-i\nabla_{{\bf x}_1}+e{\bf A}(1)\right]^2}{2m}
+e\Phi(1)\Big)\delta(1-1')
\end{equation}
describes free electrons coupled to an external electromagnetic field.  
In order to introduce the gradient expansion 
we write the Green function in
the mixed representation in terms of Wigner coordinates \begin{equation}
\label{fourier}
\check{G}(X,p) = \int\,\mbox{d}x e^{-ipx} \check{G}(X,x),
\end{equation}
where $X=([t_1+t_2]/2,[{\bf x}_1+{\bf x}_2]/2)$ is the center of mass coordinate 
and $x=(t_1-t_2,{\bf x}_1-{\bf x}_2)$ the relative one,
\begin{equation}
X=(T,{\bf R}),\;x=(t,{\bf r}),\;p=(\epsilon, {\bf p}),\;
px = -\epsilon t + {\bf p}\cdot{\bf r}.
\nonumber
\end{equation}
Notice that in the presence of both translational symmetry with respect to time and space, 
the dependence on $X$ drops out and convolution products as those in Eq.(\ref{kinetic0}) 
would reduce to simple products in Fourier space. 
In the presence of external fields or in non-equilibrium conditions, 
Fourier transforms, as  defined in (\ref{fourier}), of convolution products 
can be systematically expanded in powers of derivatives with respect to the center of mass coordinates. 
To leading order, the gradient expansion applied to Eq.(\ref{kinetic0}) yields
\begin{eqnarray}
-i\left[G_0^{-1}\stackrel{\otimes}{,}\check{G}\right]
&\approx&
\partial_{\epsilon}G_0^{-1}\partial_T \check{G} - \partial_TG_0^{-1}\partial_{\epsilon}\check{G}+
\nonumber\\
\label{gradient0}
&&
-\nabla_{\bf p}G_0^{-1}\cdot\nabla_{\bf R}\check{G} + \nabla_{\bf R}G_0^{-1}\cdot\nabla_{\bf p}\check{G}.
\end{eqnarray}
Notice that such an expansion is in our case justified by the assumption that
$p_F (\epsilon_F)$ is the biggest momentum (energy) scale of the problem.
On the r.h.s.~in the above both $\check{G}$ and $G_0^{-1}$ are functions of $(X,p)$, with
\begin{equation}
G_0^{-1}(X,p) = 
\epsilon - \frac{\left[{\bf p}+e{\bf A}(X)\right]^2}{2m}+e\Phi(X).
\end{equation}
Integrating the Keldysh component of Eq.~(\ref{gradient0}) over the energy $\epsilon$ 
leads to the l.h.s. of the Boltzmann equation for the distribution function
\begin{equation}
f(X,{\bf p})\equiv\frac{1}{2}
\left(
1+\int\,\frac{\mbox{d}\epsilon}{2\pi i} G^K(X,p)
\right).
\end{equation}
Note that the above defined quantity is \textit{not} gauge invariant.
For this reason if one is to obtain a result in the form of Eq.~(\ref{boltzmann0})
a shift of the whole equation must formally be performed -- i.e. one must send 
${\bf p}\rightarrow{\bf p}^*$ in Eq.~(\ref{gradient0}). 
This is done in Ref.~[\onlinecite{langreth1966}], though such a shift could also have been
performed \textit{before} the gradient expansion. The latter is the way followed
in Ref.~[\onlinecite{altshuler1978}], where the mixed representation is right from the beginning
defined in terms of the kinematic momentum
\begin{equation}
\label{shift0}
\check{G}(X,p)\rightarrow\int\,\mbox{d}x e^{-i[p+eA(X)]x}\check{G}(X,x),\;
A=(\Phi,{\bf A}).
\end{equation}
Unfortunately the simple and convenient concept of a ``shift'' does not work 
when non-Abelian gauges are considered.  
This is because the nature of the transformation (\ref{shift0}) 
is actually \textit{geometric}, a fact that manifests itself only
when dealing with non-commuting fields.  At its core lies the Wilson line 
$U_{\Gamma}(2,1)$, which is the exponential of the line integral of the gauge potential 
along the curve $\Gamma$ going from $1$ to $2$ (see [\onlinecite{peskin}])
\begin{equation}
\label{wilson0}
U_{\Gamma}(2,1)\equiv\mathcal{P}e^{-i\eta\int\,{\rm d}y A(y)}.
\end{equation} 
Here the symbol $\mathcal{P}$ stands for path-ordering along $\Gamma$,
whereas $\eta$ is a general coupling constant -- in the $U(1)$ case
it reduces to $e$.
In the spirit of the gradient expansion the integral (\ref{wilson0})
is evaluated for small values of the relative coordinate $x$, and it is thus reasonable to pick $\Gamma$
as the straight line connecting the two points.  For the $U(1)$ gauge it is seen that
\begin{equation}
\label{wilsonexp0}
U_{\Gamma}(2,1)\approx e^{-ieA(X)x}
\end{equation}
which is precisely the phase factor appearing in Eq.~(\ref{shift0}).
In other words, the ``shifting'' of Eq.~(\ref{gradient0}) should properly
be seen as the transformation
\begin{eqnarray}
&
\left[G_0^{-1}(1,1')\stackrel{\otimes}{,}\check{G}(1',2)\right] \rightarrow
&
\nonumber\\
&
U_{\Gamma}(X,1)\left[G_0^{-1}(1,1')\stackrel{\otimes}{,}\check{G}(1',2)\right]U_{\Gamma'}(2,X),
&
\label{transformation0}
\end{eqnarray}
where $\Gamma (\Gamma')$ is a straight line from $1(X)$ to $X(2)$\footnote{Since the gauge 
transformation properties do not depend on the choice of $\Gamma,\Gamma'$, we pick the simplest curve.} 
and matrix multiplication
over the internal indices between $U_{\Gamma}, U_{\Gamma'}$ and the commutator
is implied.

With this hindsight about the nature of the ``shift''
required to obtain Eq.~(\ref{boltzmann0}), it is possible to generalize the construction
to the non-Abelian case.
A general gauge transformation is defined as a local rotation of the second-quantized annihilation fermionic field $\psi$
\begin{equation}
\label{gauge0}
\psi'(1) = V(1)\psi(1),\; V(1)V^{\dagger}(1)=1.
\end{equation}
For the gauge potential one has
\begin{equation}
\eta A'(1) = V(1)\left[\eta A(1)+i\partial_1\right]V^{\dagger}(1).
\end{equation}
Notice that this is now a tensor with both real space and gauge indices
\begin{equation}
\eta A(1) = (e\Phi + \gamma{\Psi}^a t^a/2, e{\bf A} + \gamma\boldsymbol{\mathcal A}^a t^a/2),
\end{equation}
where we found convenient to separate the Abelian coupling constant $e$
from its non-Abelian counterpart $\gamma$.  The non-Abelian scalar potential
$\gamma\Psi^a t^a/2$ describes a Zeeman term -- 
i.e. of the kind ${\bf b}\cdot{\bf s}$.  Here ${\bf s}$ is the spin of the carriers,
whereas ${\bf b}$ could be an applied magnetic field or, 
in a ferromagnet, the exchange field due its magnetization.
The $t^a/2$'s are the generators 
of the given symmetry group, which in the $SU(2)$ case become the
Pauli matrices, $t^a=\sigma^a, a=x,y,z$.  
In the following boldfaced quantities will indicate
vectors in real space, whereas the presence of italics will denote
a gauge structure -- which, as in the above, will sometimes be written
down explicitly.  A sum over repeated indices is always implied.  
Since a Wilson line transforms covariantly, i.e. 
$U'_{\Gamma}(2,1)=V(2)U_{\Gamma}(2,1)V^{\dagger}(1)$,
it is possible to define a Green's function
\begin{equation}
\check{\tilde{G}}(1,2)\equiv U_{\Gamma}(X,1)\check{G}(1,2)U_{\Gamma}(2,X)
\end{equation}
which is \textit{locally} covariant, i.e.
\begin{equation}
\check{\tilde{G}}'(1,2) = V(X)\check{\tilde{G}}(1,2)V^{\dagger}(X).
\end{equation}
In terms of $\tilde{G}^K$ we can define a distribution function
\begin{equation}
f(X,{\bf p})\equiv\frac{1}{2}
\left(
1+\int\,\frac{{\mbox d}\epsilon}{2\pi i}\tilde{G}^K(X,p)
\right)
\end{equation}
which will be the natural generalization of $f(X,{\bf p}^*)$ from
Eq.~(\ref{boltzmann0}).  
The procedure is then clear:
\begin{enumerate}
 \item{transform the kinetic equation according to Eq.~(\ref{transformation0});}
 \item{expand the Wilson lines (see below);}
 \item{perform a gradient expansion and write everything in terms of $\tilde{G}^K(X,p)$;}
 \item{integrate over the energy $\epsilon$ to obtain the Boltzmann equation, or
  over $\xi\equiv p^2/2m-\mu$ to end up with the Eilenberger equation.}
\end{enumerate}
Postponing the discussion of the last point to the next Section,
we now consider the general expression
\begin{eqnarray}
G_0^{-1}(X,p) &=& \epsilon - H(X,p)
\nonumber\\
&=&
\epsilon - \frac{\left[{\bf p}+e{\bf A}(X)+\gamma\boldsymbol{\mathcal A}^a(X)t^a/2\right]^2}{2m}+
\nonumber\\
&&
+e\Phi(X)+\gamma\Psi^a(X)t^a/2.
\end{eqnarray}
In the Rashba model, Eq.~(\ref{rashba0}), one for example identifies
\begin{equation}
t^a=\sigma^a,\quad
\boldsymbol{\mathcal A}^a = \boldsymbol{\mathcal A}^a_R.
\end{equation} 
The procedure outlined above (points 1.-3.) 
leads to a \textit{locally} covariant equation for $\check{\tilde G}$ 
accurate to order $[(\partial_X\partial_p)(A\partial_p),(A\partial_p)^2]$ 
(see 
\footnote{This corresponds to a ``next-to-leading-order'' expansion and is necessary to couple
Abelian and non-Abelian fields to one another.  In the physically more transparent language of spin-orbit
coupling, the terms order $[(\partial_X\partial_p),(A\partial_p)],(A\partial_p)^2$
are those that produce spin-charge coupling -- and thus lead to, say, the spin Hall effect or the 
anomalous Hall effect.}):
in the mixed representation language we have formally two expansion parameters,
$\partial_X\partial_p\ll1$ -- the standard gradient expansion one -- and
$A\partial_p\ll1$ -- coming from the gauge fields.  In the $SU(2)$ case the latter
corresponds to the physical assumption that the spin-orbit energy be small
compared to the Fermi one, $\Delta_{so}/\epsilon_F\ll1$.
Even though our treatment is valid for any non-Abelian gauge, 
we now pick the $SU(2)$ gauge for definiteness' sake.  
In this case steps 1.-3. lead to 
\begin{eqnarray}
\Big(
\tilde{\partial}_T + \frac{\bf p}{m}\cdot\tilde{\nabla}_{\bf R} -
\frac{\bf p}{2m}\cdot
\left\{
\left[e{\bf E} + \gamma\boldsymbol{\mathcal E}\right]\partial_{\epsilon}, \;.\;
\right\}
+
&&
\nonumber\\
\label{covariant0}
+
\frac{1}{2}
\left\{
\boldsymbol{\mathcal F}\cdot\nabla_{\bf p}, \;.\;
\right\}
\Big)
\check{\tilde G} 
&=& 0,
\end{eqnarray}
where the symbol $\left\{\,.\,,\,.\,\right\}$ denotes the anticommutator.
The covariant (``wavy'') derivatives are
\begin{equation}
\label{covder}
\tilde{\partial}_T = \partial_T -i\gamma\left[\Psi,\;.\;\right],\;
\tilde{\nabla}_{\bf R} = \nabla_{\bf R} +i\gamma\left[\boldsymbol{\mathcal A},\;.\;\right],
\end{equation}
whereas the generalized Lorentz force reads
\begin{eqnarray}
\label{superlorentz}
\boldsymbol{\mathcal F} = 
-
\underbrace{
e\left[
{\bf E}+\frac{{\bf p}\wedge{\bf B}}{m}
\right]}_{U(1)}
-
\underbrace{
\gamma\left[
\boldsymbol{\mathcal E}
+\frac{{\bf p}\wedge\boldsymbol{\mathcal B}}{m}
\right]}_{SU(2)}.
\end{eqnarray}
The fields are given as usual in terms of the field tensor $\mathbb{F}$, 
but this has now an $SU(2)\times U(1)$ structure 
\begin{eqnarray}
&&
E_i = \mathbb{F}_{0i}^0 = -\partial_T A_i -\nabla_{R_i}\Phi
\\
&&
{\mathcal E}_i^a = \mathbb{F}_{0i}^a = -\partial_T {\mathcal A}_i^a - \nabla_{R_i}\Psi^a
 + i\gamma\left[\Psi,{\mathcal A}_i\right]^a
\\
&&
B_i = \frac{1}{2}\epsilon_{ijk}\mathbb{F}_{jk}^0,\;  
{\mathcal B}_i^a = \frac{1}{2}\epsilon_{ijk}\mathbb{F}_{jk}^a
\\
&&
\mathbb{F}_{jk}^0 = \nabla_{R_j}A_k - \nabla_{R_k}A_j
\\
&& 
\mathbb{F}_{jk}^a = \nabla_{R_j}{\mathcal A}_k^a - \nabla_{R_k}{\mathcal A}_j^a + 
i\gamma\left[{\mathcal A}_j, {\mathcal A}_k \right]^a.
\end{eqnarray}
Note that in order to obtain Eq.~(\ref{covariant0}) it is sufficient to expand the Wilson lines
to first oder in $x$, e.g.
\begin{equation}
U_{\Gamma}(X,1) \approx 1 + \eta A\frac{ix}{2}.
\end{equation}
This is not true for a general convolution of the kind $[F(1,1')\stackrel{\otimes}{,}G(1',3)]$,
with $F(1,1')$ a function with a more complicated structure than that of
$G_0^{-1}(1,1')$.  Such a case would require a second order expansion, e.g.
\begin{equation}
U_{\Gamma}(X,1) \approx 1 + \eta A\frac{ix}{2} + \eta \partial_XA\frac{ix^2}{8} - \eta^2A^2\frac{x^2}{8}, 
\end{equation}
and would lead to a rather more complicated equation. 

To complete our preparatory work for the derivation of the kinetic Boltzmann or Eilenberger equations, 
we need to introduce the effect of disorder.
Within the Keldysh formalism this is done by the addition of a 
self-energy contribution on the r.h.s. of  Eq.~(\ref{kinetic0})
\begin{equation}
-i\left[\check{\Sigma}(1,2)\stackrel{\otimes}{,}\check{G}(2,3)\right],
\end{equation}  
which can be manipulated just as the ``free'' ($G_0^{-1}$) term. 
In spin-orbit coupled systems the presence of disorder can have a number of interesting
effects.  Indeed, phenomena like the spin Hall effect, anomalous Hall effect or related ones can have
both an \textit{intrinsic} and an \textit{extrinsic} origin
\cite{engel2007, sinitsyn2008b, inoue2009}.
This depends on whether they arise from fields due
to the band or device structure, or from those
generated by impurities.
In the latter case skew-scattering and side-jump contributions
to the dynamics appear \cite{dyakonov1971, hirsch1999}.  
For a discussion of these issues see \cite{nozieres1973,engel2005,
sinitsyn2008b, hankiewicz2008, raimondi2009}.
In the following we limit ourselves to the treatment
of \textit{intrinsic} effects in the presence of spin independent
disorder. 
We consider elastic scattering with probability $W=W({\bf p}-{\bf p}')$ and  quasiparticle lifetime 
$\tau^{-1}=2\pi \sum_{{\bf p'}}\delta (\epsilon_{\bf p}-\epsilon_{\bf p'}) W({\bf p}-{\bf p}')$. 
In the Born approximation, the disorder self-energy in the mixed representation reads
\begin{equation}
\label{noncovsigma}
\check{\Sigma}(X,{\bf p},\epsilon) = \sum_{{\bf p}'}W({\bf p}-{\bf p}')\check{G}(X,{\bf p}',\epsilon).
\end{equation}
From Eq.~(\ref{noncovsigma}) one obtains that the locally covariant self-energy $\check{\tilde{\Sigma}}$ is
\begin{equation}
\check{\tilde{\Sigma}}(X,{\bf p},\epsilon) = \sum_{{\bf p}'}W({\bf p}-{\bf p}')
\check{\tilde{G}}(X,{\bf p}',\epsilon), 
\end{equation}
which in turn implies 
\begin{equation}
\label{covariantkernel0}
\tilde{I}[\tilde{G}] = -i\left[\check{\tilde{\Sigma}},\check{\tilde{G}}\right].
\end{equation}
Note that for a leading order description of the coupling between spin [$SU(2)$] and charge [$U(1)$],
corrections $\mathcal{O}(A\partial_p)$ in the collision integral are enough.
Notice also that, whereas $\check{G}$
is peaked at the different folds of the spin-split Fermi surface,
the peaks of $\check{\tilde{G}}$ are ``shifted'' and thus located
on the Fermi surface in the absence of spin-orbit coupling.


\section{The Boltzmann and Eilenberger equations}
\label{secintegration}

The question of whether to integrate the locally covariant kinetic equation with
respect to $\epsilon$ or to $\xi=p^2/2m - \mu$ depends on the physical situation.
If the spectral density $i(G^R-G^A)$ is not ``$\delta$-like'' as a function of $\epsilon$
the energy integration is formally impracticable.  The $\xi$-integration
on the other hand is capable of justifying a Boltzmann-like approach
even when the first approach fails or looks severely limited \cite{rammer1986, prange1964}.
For the case considered of a degenerate gas of free electrons
colliding elastically with impurities both procedures are viable,
provided the condition $\epsilon_F\tau\gg1$ holds -- since, as mentioned
before, all quantities appearing in Eq.~(\ref{covariantkernel0})
are peaked at the $\Delta_{so}=0$ Fermi surface.

\subsection{Boltzmann ($\epsilon$-integration)}

Energy integration of the Keldysh component of Eqs.~(\ref{covariant0}) and (\ref{covariantkernel0})
yields a Boltzmann-like kinetic equation for the $2\times 2$ 
matrix distribution function $f(X,{\bf p})$
\begin{equation}
\label{covariant1}
\left(
\tilde{\partial}_T + \frac{{\bf p}}{m}\cdot\tilde{\nabla}_{\bf R} 
+\frac{1}{2}\left\{\boldsymbol{\mathcal F}\cdot
\nabla_{\bf p}, \;.\; \right\}
\right)f(X,{\bf p})=\tilde{I}[f] 
\end{equation}
with the covariant derivatives and the generalized Lorentz 
force $\boldsymbol{\mathcal F}$ defined respectively 
as in Eqs.~(\ref{covder}) and (\ref{superlorentz}), 
and where the collision integral reads
\begin{equation}
\label{collision0}
\tilde{I}[f]
=
-2\pi\sum_{{\bf p}'}W({\bf p}-{\bf p}')\delta(\epsilon_{\bf p}-\epsilon_{{\bf p}'})
\left[
f(X,{\bf p})-f(X,{\bf p}')
\right].
\end{equation}
Notice that Eqs.~(\ref{covariant1})-(\ref{collision0}) are formally valid both in two and three dimensions.
However, since the physical system we have in mind is a two-dimensional electron gas,
from now on we restrict ourselves to two dimensions.
Observable properties are conveniently expressed via the matrix
density, $\rho $, 
and current, $\boldsymbol{\mathcal J}$,
\begin{eqnarray}
\rho(X) &=& \int\,\frac{{\mbox d}^2p}{(2\pi)^2}f(X,{\bf p})
\label{obs1}
\\
\label{current0}
\boldsymbol{\mathcal J}(X) &=& 
\int\,\frac{{\mbox d}^2p}{(2\pi)^2}\frac{\bf p}{m}f(X,{\bf p}),
\end{eqnarray}
which obey the following generalized continuity equation,
\begin{equation}
\label{continuity}
\tilde{\partial}_T\rho(X) + \tilde{\nabla}_{\bf R}\cdot\boldsymbol{\mathcal J}(X) = 0,
\end{equation}
derived by integrating Eq.~(\ref{covariant1}) over the momentum.
Observables like the the particle and spin densities, $n$ and $s^a$,
and the particle and spin currents, ${\bf j}^{0}$ and ${\bf j}^a$,
can be evaluated as 
\begin{eqnarray}
n(X) &=& {\rm Tr}\left[\rho(X)\right]
\\
s^a(X) &=& \frac{1}{2}
          {\rm Tr}\left[\sigma^a\rho(X)\right]
\\
{\bf j}^0(X) &=& {\rm Tr}\left[\boldsymbol{\mathcal J}\right]
\\
{\bf j}^a(X) &=& \frac{1}{2}{\rm Tr}\left[\sigma^a\boldsymbol{\mathcal J}\right].
\end{eqnarray}
One can check that these expressions agree with their microscopic
definitions \cite{jin2006, tokatly2008}.

Eq.~(\ref{covariant1}) is the first main result of the paper.  Though the idea of rewriting
spin-orbit interaction in terms of non-Abelian gauge fields is no novelty, 
we are not aware of a Boltzmann formulation in the above form.
Whereas in Refs.~[\onlinecite{khaetskii2006},\onlinecite{trushin2007}] the collision integral and
the velocity are non-diagonal in the charge-spin indices, 
here their structure is simpler.  The gauge fields appear only in
the covariant derivatives, describing precession of the spins around
the external magnetic field and the internal spin-orbit one,
and in the generalized Lorentz force, which couples the spin
and charge channels.

\subsection{Eilenberger ($\xi$-integration)}
 
The integration over $\xi$ of Eqs.~(\ref{covariant0}) and (\ref{covariantkernel0})
yields in two dimensions the Eilenberger equation
\begin{eqnarray}
\Big(
\tilde{\partial}_T + v_F\hat{\bf p}\cdot\tilde{\nabla}_{\bf R}
-\frac{1}{2}\partial_{\epsilon}\left\{
\left[\frac{{\bf p}(\epsilon)}{m}\cdot(e{\bf E}+\gamma\boldsymbol{\mathcal E})\right],\;.\;
\right\}
+
&&
\nonumber
\\
+\frac{1}{2p_F}
\left\{
\boldsymbol{\mathcal F}(p_F,\varphi)\cdot
\left[-\hat{\bf p}+
\hat{\boldsymbol\varphi}\partial_{\varphi}
\right],\;.\;
\right\}
\Big)\tilde{g}^K
&&
\nonumber\\
\label{covariant2}
=
-2\pi N_0\int\,\frac{{\mbox d}\varphi'}{2\pi}W(\varphi-\varphi')
\left[\tilde{g}^K(\varphi)-\tilde{g}^K(\varphi')\right],
&&
\end{eqnarray}
where $\hat{\bf p}=(\cos\varphi,\sin{\varphi}),
\hat{\boldsymbol\varphi}=(-\sin\varphi,\cos\varphi)$,
$W(\varphi-\varphi')$ is the scattering amplitude at the Fermi surface,
and $\tilde{g}^K$ is the Keldysh component of the covariant quasiclassical Green's function
\begin{equation}
\check{\tilde{g}}(X,\varphi,\epsilon)\equiv
\frac{i}{\pi}\int\,{\mbox d}\xi\check{\tilde{G}}(X,\varphi,\epsilon,\xi).
\end{equation}
Notice that the energy derivative $\partial_{\epsilon}$ acts
on the whole anticommutator, i.e. on $\tilde{g}^K$ too.
Just as in the Boltzmann case, and as opposed to what happens in the literature
\cite{shytov2006, raimondi2006},
the velocity and the collision integral have here a simple diagonal structure, 
whereas the gauge fields appear only in the covariant derivatives and force terms.
The collision integral will be extensively discussed in Appendix~\ref{appendixcollision}
to make an explicit comparison with Ref.~[\onlinecite{shytov2006}] possible.  
Integration of Eq.~(\ref{covariant2}) over the energy and the angle
leads again to the continuity equation (\ref{continuity}), this time
with densities and currents expressed in terms of $\tilde{g}$
\begin{eqnarray}
\rho(X) &=& -\frac{N_0}{2}\int\,{\mbox d}\epsilon\langle\tilde{g}^K\rangle\label{densitycovariant}
\\
\boldsymbol{\mathcal J}(X) &=& 
-\frac{N_0}{2}\int\,{\mbox d}\epsilon v_F\langle \hat{\bf p}\tilde{g}^K\rangle,
\end{eqnarray}
where $\langle...\rangle$ denotes the angular average.
Recall that when expressing physical quantities in terms
of the standard quasiclassical Green's function
$\check{g}=i/\pi\int\,{\mbox d}\xi\check{G}$ equilibrium \textit{high-energy} contributions
are missed \cite{rammer1986, schwab2003}.  
For instance, Eq.(\ref{densitycovariant}) for the particle density when only $U(1)$ fields are present
would be written, in terms of $g^K$, as
\begin{equation}
\label{densitynoncovariant}
\rho(X) = -\frac{N_0}{2}\int\,{\mbox d}\epsilon\langle g^K\rangle+N_0 e \Phi (X),
\end{equation}
with the second term due to the scalar potential originating from the high-energy part.
A virtue of the present formulation  is that
such contributions are by construction included
in the covariant $\tilde{g}$.  
Moreover notice that, whereas in the presence of spin-orbit coupling
the usual normalization condition $\check{g}^2=1$ is modified and
becomes momentum-dependent\cite{raimondi2006}, in the covariant formulation
$\check{\tilde{g}}^2=1$ holds -- see Appendix~\ref{tildenormal}.
The normalization condition is established by direct calculation ``at infinity'',
i.e. where, far from the perturbed region, the Green's function
reduces to its equilibrium form.  It plays the role of a boundary condition
imposed on Eq.~(\ref{covariant2}), and thus defines its solution uniquely\cite{eckern1981, shelankov1985}.
In the presence of interfaces between different regions wave functions have
to be matched, and this can be translated into a
condition to be fulfilled by the quasiclassical Green's function on either side
of the interfaces\cite{zaitsev1984,millis1988}.  Recently some very general such boundary
conditions for multiband systems were obtained\cite{eschrig2009},
though valid only as long as the spin and charge channels are decoupled.
When this is not anymore the case, things are complicated by the momentum dependence of
the normalization and, as far as we are aware of, beyond the present treatment of boundaries.
The covariant formulation in terms of $\tilde{g}$ suggests however the possibility
for a non-trivial extension of the known boundary conditions 
to the case in which spin and charge channels are coupled, 
precisely because of the simple normalization of $\tilde{g}$.


\section{The diffusive regime}
\label{secdiff}

Our goal in this Section is the discussion of spin-charge coupled dynamics
in the diffusive regime.  
Formally, the ``non-Abelian'' Boltzmann equation (\ref{covariant1}) can
be solved just as in the $U(1)$ case.
We expand the angular dependence of  
the distribution $f$ in harmonics,
$f=\langle f\rangle+2\hat{\bf p}\cdot{\bf f}+...$
and use the expansion in Eq.~(\ref{covariant1}) to obtain an explicit expression
for ${\bf f}$,
\begin{eqnarray}
{\bf f} 
&\approx&
-\frac{\tau_{tr}p}{2m} \tilde \nabla_{\bf R}\langle f\rangle 
-
\frac{\tau_{tr}}{2}\langle\hat{\bf p}
\left\{
\boldsymbol{\mathcal F}\cdot\nabla_{\bf p},\langle f\rangle
\right\}\rangle+
\nonumber\\
&&
-
\frac{\tau_{tr}}{2}\langle\hat{\bf p}
\left\{
\boldsymbol{\mathcal F}\cdot\nabla_{\bf p},
\left(2\hat{\bf p}\cdot{\bf f}\right)
\right\}\rangle
\nonumber\\
\label{distribexp1}
&\equiv&
{\bf f}_{\rm diff} 
+ 
{\bf f}_{\rm drift}
+
{\bf f}_{\rm Hall}.
\end{eqnarray}
In the above $\tau_{tr}$ is the usual transport time
\begin{equation}
\frac{1}{\tau_{tr}} = 2\pi N_0\int\,\frac{{\mbox d}\varphi'}{2\pi}W(\varphi-\varphi')
[1-\cos(\varphi-\varphi')],
\end{equation}
which depends on the energy $\xi = \epsilon_p - \epsilon_F $ through the scattering 
probability $W(\varphi-\varphi')=W(\xi,\xi';\varphi-\varphi')|_{\xi=\xi'}$.
The diffusion term ${\bf f}_{\rm diff}$ is related to the (covariant) 
derivative of the angular average of the distribution
function, i.e. to the derivative of the charge and spin densities.
The drift term ${\bf f}_{\rm drift}$ arises from the second term on the r.h.s. of Eq.~(\ref{distribexp1}),
in which only the ``electric'' part of the Lorentz force, 
i.e. $-[e{\bf E}+\gamma\boldsymbol{\mathcal E}]$, contributes.
The Hall component ${\bf f}_{\rm Hall}$ comes instead from  the third term on the r.h.s. of
Eq.~(\ref{distribexp1}), and is due to the ``magnetic'' part of the Lorentz force,
$-{\bf p}\wedge\left[e{\bf B}+\gamma\boldsymbol{\mathcal B}\right]/m$.
Using Eq.~(\ref{distribexp1}) into Eq.~(\ref{current0}) one finally has
\begin{eqnarray}
\boldsymbol{\mathcal J} 
&=&
\int\,\frac{{\mbox d}^2p}{(2\pi)^2}\frac{\bf p}{m}
\left[
\langle f\rangle + 2\hat{\bf p}\cdot{\bf f}+\dots
\right]
\nonumber\\
&\approx&
\int\,\frac{{\mbox d}^2p}{(2\pi)^2}\frac{\bf p}{m}
2\hat{\bf p}\cdot\left[
{\bf f}_{\rm diff} 
+ 
{\bf f}_{\rm drift}
+
{\bf f}_{\rm Hall}
\right]
\nonumber\\
\label{protocurrent}
&\equiv&
\boldsymbol{\mathcal J}_{\rm diffusion}+\boldsymbol{\mathcal J}_{\rm drift}+
\boldsymbol{\mathcal J}_{\rm Hall}.
\end{eqnarray}
The drift  current is straightforwardly computed
\begin{eqnarray}
\label{drift}
\boldsymbol{\mathcal J}_{\rm drift} &=& \frac{1}{2}\left\{ \sigma(\mu),
e{\bf E}+\gamma\boldsymbol{\mathcal E}\right\}, \, \, \sigma(\mu) = -N_0 D(\mu)
\end{eqnarray}
where $N_0$ is the density of states, $D(\mu)$ the
energy dependent diffusion constant and $\mu$ the (spin dependent)
electrochemical potential.
Since we assume fields that are small compared to the Fermi energy, 
it is often sufficient to replace $D(\mu)$
by its value at the Fermi energy and in the absence of the $U(1)$ and $SU(2)$ fields. 
In the examples we discuss it will be important to go one step beyond this simple approximation, 
for which we obtain
\begin{equation} \label{eq55}
D(\mu) \approx D(\epsilon_F) + \partial_\xi D \,  ( \rho - N_0 \epsilon_F)/N_0
\end{equation}
with 
\begin{eqnarray}
&&
D(\epsilon_F) = \frac{v^2_F\tau_{tr}}{2},\quad\partial_{\xi}D=\frac{\tau_{tr}}{m}(1+\gamma_0/2),
\\
&&
\gamma_0/2 = \frac{m v_F^2 }{\tau_{tr} } \partial_{\xi} \tau_{tr}.
\end{eqnarray}
The factor $\gamma_0$ is defined to make direct contact to
Ref.~[\onlinecite{engel2007b}].
Notice that due to the expansion in (\ref{eq55}) the diffusion constant $D(\mu)$ 
becomes a spin dependent object,
\begin{equation}
D(\mu) = D^0 + D^a \sigma^a
\end{equation}
with $D^0 \approx D(\epsilon_F)$ and $D^a \approx \partial_{\xi } D s^a/N_0$.

The calculation of the diffusion current is slightly more involved: the momentum
integration is delicate, since the integrand has a nontrivial matrix structure
and is out of equilibrium.  In order to ``extract'' such a structure we first write
\begin{equation}
\label{diffusioncurrent1}
\boldsymbol{\mathcal J}_{\rm diffusion}\equiv -\frac{1}{2}
\left\{
{\mathcal D},\tilde{\nabla}\rho
\right\},
\end{equation}
thus defining a diffusion constant ${\mathcal D}$ which is now a matrix.
Extending the Einstein relation to the present non-Abelian case
will give ${\mathcal D}$ an explicit form.
At equilibrium one has
\begin{equation}
\rho_{eq} = N_0 [e\Phi + \gamma\Psi]+ N_0 \epsilon_F ,
\end{equation}
and as -- again, at equilibrium -- the diffusion current balances out the drift one
\begin{eqnarray} 
\boldsymbol{\mathcal J}_{\rm drift} 
&=& 
-\boldsymbol{\mathcal J}_{\rm diffusion}
\nonumber\\
&=&
\frac{1}{2}
\left\{
{\mathcal D},\tilde{\nabla}\rho_{eq}
\right\}
\nonumber\\
&=&
-\frac{1}{2}
\left\{
{\mathcal D},N_0[e{\bf E} + \gamma\boldsymbol{\mathcal E}]
\right\}
\end{eqnarray}
there follows
\begin{equation}
{\mathcal D} = D(\mu)
\end{equation}
as to be expected.

The Hall term ${\bf f}_{\rm Hall}$ can be obtained from the equation implicit in Eq. (\ref{distribexp1})
\begin{equation}
\label{hall_drift}
{\bf f}_{\rm hall } =  \frac{\tau_{tr}}{2m}
\left\{e{\bf B}+\gamma\boldsymbol{\mathcal B}
\stackrel{\wedge }{, }
{\bf f}\right\}
,  \end{equation}
from which we find
\begin{equation}
\boldsymbol {\mathcal J}_{\rm Hall }  =   
 \frac{1 }{4 m }
\left\{
  e {\bf B} + \gamma \boldsymbol {\mathcal B} 
  \stackrel{ \wedge }{ , }  
  \{ \tau_{tr}(\mu) , \boldsymbol {\mathcal J}  \}
\right\}
,\end{equation}
with
\begin{eqnarray}
 \tau_{tr}(\mu)  & =  &\tau_{tr} + \partial_{\xi} \tau_{tr} (\rho - N_0 \epsilon_F)/N_0 \\
                 & =  &\tau_{tr} + \frac{\gamma_0 }{2} \frac{\tau_{tr}}{m v_F^2 } \frac{\rho - N_0 \epsilon_F }{N_0} 
. \end{eqnarray}

To be more explicit we give the expressions for the particle current ${\bf j}^0$ and spin current ${\bf j}^a$
\begin{eqnarray}
{\bf j}^0 &=& -D \Big(\nabla n +2eN_0{\bf E}\Big) - 2 D^{a}\Big([\tilde{\nabla}s]^a+\frac{\gamma N_0}{2}
\boldsymbol{\mathcal E}^a\Big)+
\nonumber\\
\label{current1}
&&
-\frac{e\tau_{tr}}{m}{\bf j}^0\wedge{\bf B} -\frac{\gamma{ \tau}_{tr}}{m}{\bf j}^a\wedge\boldsymbol{\mathcal B}^a
\\
{\bf j}^a &=& -\frac{1}{2} D^{a}\Big(\nabla n +2eN_0{\bf E}\Big) - D \Big([\tilde{\nabla}s]^a+\frac{\gamma N_0}{2}
\boldsymbol{\mathcal E}^a\Big)+
\nonumber\\
\label{current2}
&&
-\frac{e{\tau}_{tr}}{m}{\bf j}^a\wedge{\bf B} -\frac{\gamma\tau_{tr}}{4m}{\bf j}^0\wedge\boldsymbol{\mathcal B}^a
.\end{eqnarray}
Here we included the Hall current only in the leading approximation, i.e. $\tau_{tr} (\mu) \approx \tau_{tr}(\epsilon_F)$.
The diffusion equations for charge and spin are obtained by inserting 
Eqs.~(\ref{current1}) and (\ref{current2}) into the continuity equation 
(\ref{continuity}).


\section{Two examples}
\label{secexamples}

\subsection{Effect of an in-plane magnetic field}

As a first simple example that shows how the formalism works, we obtain
and solve the Bloch equations for a Rashba 2DEG driven by an electric field along $x$ 
and in the presence of an in-plane Zeeman field along $x$.
This is the same geometry considered in Refs.~[\onlinecite{raimondi2009, milletari2008, engel2007b}].

The $U(1)$ fields read
\begin{equation}
{\bf E} = (E,0,0),\quad {\bf B}=0
\end{equation}
while, since the Zeeman field enters the Hamiltonian through the scalar potential 
$\gamma\Psi^x\sigma^x/2\equiv b^x\sigma^x/2$,
the $SU(2)$ ones are 
\begin{equation}
\gamma\boldsymbol{\mathcal E} = 2m\alpha b^x(\sigma^z/2,0,0),\quad
\gamma\boldsymbol{\mathcal B} = -(2m\alpha)^2(0,0,\sigma^z/2).
\end{equation}
From the expressions for the currents derived in the previous section, one obtains
in the homogeneous limit a set of Bloch equations which generalizes those
appearing in Ref.~[\onlinecite{raimondi2009}] 
to the case of angle-dependent scattering -- but in the absence of extrinsic effects --,
namely
\begin{eqnarray}
\dot{\bf s} &=&
-\hat{\Gamma}\left[{\bf s}-{\bf b}N_0/2+e\alpha\tau_{tr}N_0\hat{\bf z}\wedge{\bf E}\right]+
\nonumber\\
\label{bloch}
&&
-\left[{\bf b}-2e\alpha\tau_{tr}(1+\gamma_0/2)\hat{\bf z}\wedge{\bf E}\right]\wedge{\bf s}.
\end{eqnarray}
Here $\hat{\Gamma}=1/\tau_{DP}\,{\rm diag}(1,1,2)$ is the
relaxation matrix, with $1/\tau_{DP}=(2m\alpha)^2D$ the Dyakonov-Perel relaxation rate.
Notice that the electric field in the first and in the second term
on the r.h.s. of Eq. (71) has a different origin. While the first
term is traced back to the (spin) Hall current and therefore to the
SU(2) magnetic field, the second term can be traced back to the drift
current.
The factor $\gamma_0$ that appears due to the energy dependence of
the scattering time has an important impact on the static solution
of the Bloch equations.
When $\gamma_0 =0 $ we find
\begin{equation}
{ \bf s} = {\bf b}N_0/2 - e\alpha\tau_{tr} N_0 \hat{\bf z}\wedge{\bf E}
\end{equation}
i.e. the effects of the Zeeman and the electric field on the spin
polarization are simply additive. This is not anymore the case if
$\gamma_0 \neq 0$, in which case we find in the limit of weak electric and
magnetic fields (in our geometry both in $x$-direction)
\begin{equation}
\label{spins}
s^x=s^x_{eq},\quad s^y=-e\alpha\tau_{tr}N_0E,\quad 
s^z=-\gamma_0\frac{e\alpha\tau_{tr}N_0Eb^x}{4(2m\alpha)^2D},
\end{equation}
that is, in-plane fields generate an out-of-plane 
spin polarization\cite{engel2007b}
\footnote{Notice that the above does not agree with the results 
found in Ref.~[\onlinecite{milletari2008}], though 
the discrepancy is no surprise: in [\onlinecite{milletari2008}]
a different model of disorder was used, one in which the 
dependence of the scattering amplitude on the modulus of the momentum
was neglected.}.
In the above $s^x_{eq}=b^xN_0/2$.

\subsection{Charge current from time dependent spin-orbit coupling}

\begin{figure}
\includegraphics[width=.45\textwidth]{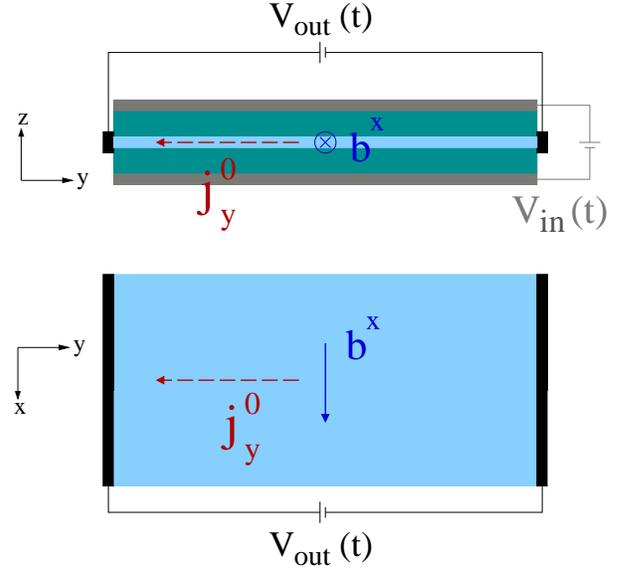}
\caption{(color online) The Rashba spin-orbit coupling constant $\alpha$ is made time
dependent by applying a time dependent gate potential $V_{in}(t)$.
The light (blue) area represents a two-dimensional electron gas inside a heterostructure.
When an in-plane magnetic field along $x$, $b^x$, is also switched on, a charge current
$j^0_y$ flowing along $y$ and proportional to $\dot{\alpha}$ is generated, 
its actual sign depending on the sign of $\dot{\alpha}$.
The induced voltage drop in the transverse direction $V_{out}(t)$
can then be used to measure the strenght of the Rashba interaction.}
\label{theonlyfig}
\end{figure}
The Rashba spin-orbit coupling constant $\alpha$ arises from the potential
confining the 2DEG and is thus tunable by a gate voltage: if the latter is time dependent, so is the former.
Let us then consider the Rashba Hamiltonian for a time-dependent
Rashba parameter, $\alpha\rightarrow\alpha(T)$.
In the non-Abelian language this means that the $SU(2)$ vector potential becomes
time dependent,
and therefore a spin dependent electric field is generated.  
Explicitly we have
\begin{eqnarray}
\gamma\boldsymbol{\mathcal E} &=&
2m\dot{\alpha}\left(\sigma^y/2,-\sigma^x/2,0\right)\label{e_field}
\\
\gamma\boldsymbol{\mathcal B} &=& -\left(2m\alpha\right)^2(0,0,\sigma^z/2),\label{b_field}
\end{eqnarray}
with $\dot{\alpha}=\partial_T\alpha$.
The $SU(2)$ electric field leads to the appearance of in-plane spin currents,
as discussed in Ref.~[\onlinecite{tang2005}].
However it does not generate a charge
current, since it acts with opposite sign on particles with different
spin: the net field obtained after averaging over all
particles is zero. This is not anymore the case if a magnetic field is also present.
Say the latter points in
$x$-direction, then a nonzero average electric field in the $y$ direction appears,
given by 
$\gamma \langle {\mathcal E}_y \rangle = - m \dot \alpha
\langle \sigma_x \rangle  = - 2 m \dot \alpha s^x/n $ (here $\langle...\rangle$ denotes the average
over all particles).  We then expect
a particle current in $y$-direction of the order $j_y = - 2 D N_0
\gamma \langle {\mathcal E}_y \rangle$. 
We now make the argument quantitative.
Let us apply an in-plane Zeeman field along $x$, as shown in Fig.~\ref{theonlyfig}.
Then the $SU(2)$ electric and magnetic fields are
\begin{eqnarray}
\gamma\boldsymbol{\mathcal E} &=&
2m\left(\dot{\alpha}\sigma^y/2+\alpha b^x\sigma^z/2,-\dot{\alpha}\sigma^x/2,0\right),
\\
\gamma\boldsymbol{\mathcal B} &=& -\left(2m\alpha\right)^2(0,0,\sigma^z/2).
\end{eqnarray} 
Note that the structure of the Bloch equations (\ref{bloch}) is
\textit{not} modified, so that
the stationary spin density is
\begin{equation}
s^x = s^x_{eq},\quad s^y = 0,\quad s^z = 0,
\end{equation}
with $s^x_{eq}=N_0b^x/2$ as before. As expected, the $SU(2)$ electric field generates a
particle current flowing along $y$,
\begin{eqnarray}
j^0_y &=& -2D^x\frac{N_0}{2}\gamma{\mathcal E}^x_y
\nonumber\\
&=&
\tau_{tr}\dot{\alpha}N_0b^x(1+\gamma_0/2), \label{inverse_edel}
\end{eqnarray}
having used $D^x = \tau_{tr}/m(1+\gamma_0/2)s^x$.
Finally, for a general direction of the in-plane magnetic field ${\bf b}$ the charge current is given by
\begin{equation}
\label{current3}
{\bf j}^0 = \tau_{tr}N_0\dot{\alpha}(1+\gamma_0/2)\hat{\bf z}\wedge{\bf b}.
\end{equation}
Such an effect could provide an alternative way of estimating the strenght
of the Rashba interaction, since other spin-orbit mechanisms would not gain
any time dependency from a modulated confining potential.

%
%
%
%
%
%


\section{Conclusions}

We showed how to microscopically derive the generalized Boltzmann and Eilenberger equations
in the presence of non-Abelian gauge fields.
In the $SU(2)$ case such equations can be used to describe spin-charge coupled
dynamics in two-dimensional systems whose Hamiltonians include linear-in-momentum 
spin-orbit coupling terms.  All degrees of freedom are treated symmetrically
and the proper identification of the physical quantities follows naturally from the form of the continuity
equation.  Considering elastic disorder, we obtained
results which hold as long as $\epsilon\gg1/\tau,\Delta_{so}$ and for arbitrary values of $\Delta_{so}\tau$.
In particular, we showed that by using the covariant quasiclassical Green function, 
the collision integral in the kinetic equation is not affected by the gauge fields, 
which only appear to modify the hydrodynamic derivative.
We expect that this nice disentanglement of gauge fields and disorder effects 
in the Boltzmann and Eilenberger equations may prove very useful 
when considering quantum corrections\cite{chalaev2005,chalaev2009}. 
We also expect the approach to allow
for a generalization of the boundary conditions for the Eilenberger equation
to the case in which spin and charge channels are coupled.
When discussing the diffusive regime, we first obtained Bloch-like equations for
the spin and charge, and then exploited them to predict a novel effect.
Finally, we note that by making the non-Abelian coupling constant
momentum dependent, $\gamma\rightarrow\gamma(\bf p)$, it may be possible to
extend the present formalism to include Hamiltonians
with more general forms of spin-orbit interaction -- i.e. not
limited to being linear in momentum. 

We thank U. Eckern, J. Rammer and M. Dzierzawa for discussions. 
This work was supported by the Deutsche Forschungsgemeinschaft
through SFB 484 and SPP 1285 and partially supported by
EU through PITN-GA-2009-234970.


\appendix
\section{Quasiclassical normalization condition and the collision integral}
\label{tildenormal}

Consider a quantity $F(1,2)$ which is non-locally covariant, i.e. which 
under the gauge transformation Eq.~(\ref{gauge0}) transforms according to $F'(1,2)=V(1)F(1,2)V^{\dagger}(2)$.
Its locally covariant counterpart reads $\tilde{F}(1,2)\equiv U_{\Gamma}(X,1)F(1,2)U_{\Gamma}(2,X)$,
where $X=([t_1+t_2]/2,[{\bf x}_1+{\bf x}_2]/2)$.  
In Wigner coordinates, up to $\mathcal{O}(A\partial_p)$ accuracy, one has
\begin{equation}
\label{prexi}
\tilde{F} = F - \frac{1}{2}\left\{A\partial_p,F \right\}.
\end{equation}
We define the $\xi$-integrated functions
\begin{eqnarray}
\label{defnormal}
f(\epsilon, \varphi, X) &=& \frac{i}{\pi}\int\,{\rm d}\xi F({\bf p},\epsilon, X),
\\
\label{deftilde}
\tilde{f}(\epsilon, \varphi, X) &=& \frac{i}{\pi}\int\,{\rm d}\xi \tilde{F}({\bf p},\epsilon, X).
\end{eqnarray}
Let us start by assuming for simplicity 
$A=(0, \boldsymbol{\mathcal A}^a\sigma^a/2)$ -- that is, we have neither
electric nor magnetic fields, only spin-orbit coupling -- and setting the $SU(2)$
coupling constant to one, $\gamma=1$.  Moreover, the functions $F,\tilde{F}$
are assumed to be peaked at the Fermi surface $\xi=0$ or in its vicinity.
Thus, by $\xi$-integrating Eq.~(\ref{prexi}) by parts one has
\begin{equation}
\label{postxi}
\tilde{f} = f + \frac{1}{2}
\left\{ 
\frac{\boldsymbol{\mathcal A}\cdot\hat{\bf p}}{p_F}
-
\frac{\boldsymbol{\mathcal A}\cdot\hat{\boldsymbol\varphi}}{p_F}\partial_{\varphi},
f
\right\},
\end{equation}
with $p_F$ the Fermi momentum in the absence of spin-orbit coupling.
The presence of a $U(1)$ vector potential can be handled just the same way,
whereas the inclusion of the scalar potentials $e\Phi+\Psi^a\sigma^a/2$
is trivial and amounts to a shift of the energy argument of $f$
\begin{equation}
\label{postxibis}
\tilde{f} = f -\frac{1}{2}\left\{(e\Phi+\Psi)\partial_{\epsilon}, f\right\}.
\end{equation}
Therefore, in the presence of a general 4-potential $A = (e\Phi+\Psi,e{\bf A}+
\boldsymbol{\mathcal A}^a\sigma^a/2)$, one has
\begin{eqnarray}
\tilde{f} &=& f - \frac{1}{2}\left\{(e\Phi+\Psi)\partial_{\epsilon}, f\right\}+
\nonumber\\
&&
+
\frac{1}{2}
\left\{ 
\frac{\left(e{\bf A}+\boldsymbol{\mathcal A}\right)\cdot\hat{\bf p}}{p_F}
-
\frac{\left(e{\bf A}+\boldsymbol{\mathcal A}\right)\cdot\hat{\boldsymbol\varphi}}{p_F}\partial_{\varphi},
f
\right\}.
\label{postxifinal}
\end{eqnarray}

We now use Eqs.~(\ref{prexi})-(\ref{postxifinal}) to show
\begin{itemize}
\item{how the $\xi$-integrated Green's functions $g$ and $\tilde{g}$ are related,
and what this implies for the latter's normalization;}
\item{that the collision integral Eq.~(\ref{covariantkernel0}) is equivalent to
the one appearing in [\onlinecite{shytov2006}].}
\end{itemize}   

\subsection{About $g$ and $\tilde{g}$}
Take $F=\check{G}$.
From Eq.~(\ref{postxi}) one obtains
\begin{equation}
\check{\tilde{g}} =\check{g} + \frac{1}{2}
\left\{\frac{\boldsymbol{\mathcal A}\cdot\hat{\bf p}}{p_F}
-\frac{\boldsymbol{\mathcal A}\cdot\hat{\boldsymbol\varphi}}{p_F}\partial_{\varphi},
\check{g}
\right\},
\end{equation} 
where we did not write down explicitly all the dependencies, since no confusion should arise.
Direct calculations show \cite{raimondi2006}
\begin{eqnarray}
g^{R,A} &=& \pm\left(1-\frac{\boldsymbol{\mathcal A}\cdot\hat{\bf p}}{p_F}\right)
\\
g^{K}_{eq} &=& 2\tanh\left(\epsilon/2T\right)\left(1-\frac{\boldsymbol{\mathcal A}\cdot\hat{\bf p}}{p_F}\right), 
\end{eqnarray}
$T$ being the temperature and the result for the Keldysh component being valid at equilibrium.
It follows that to order $|\boldsymbol{\mathcal A}|/p_F$ -- i.e. $\Delta_{so}/\epsilon_F$ -- the 
locally covariant $\xi$-integrated Green's function has no $SU(2)$ (spin) structure
\begin{eqnarray}
\tilde{g}^{R,A} &=& \pm1
\\
\label{distributiontilde}
\tilde{g}^{K}_{eq} &=& 2\tanh\left(\epsilon/2T\right)
\end{eqnarray}
and satisfies the standard normalization condition $\check{\tilde g}^2 = \check{1}$.
Recall this has the meaning of a boundary condition satisfied by $\check{\tilde g}$, 
and so is not affected by the introduction of driving electromagnetic [U(1)] fields 
\cite{shelankov1985} or of a Zeeman term.  Indeed, we saw that including the scalar potentials
$e\Phi$ and $\boldsymbol{\Psi}$
simply ``shifts'' the energy argument of $\tilde{g}^K_{eq}$
\begin{equation}
\tilde{g}^K_{eq} = \left[1-\frac{1}{2}\left\{(e\Phi+\Psi)\partial_{\epsilon}, \;.\; \right\}\right]
2\tanh\left(\epsilon/2T\right).
\end{equation}

\subsection{The collision integral}
\label{appendixcollision}    

Take $F=-i[\check{\Sigma},\check{G}]^K\equiv C$, and so 
$\tilde{F}=-i[\check{\tilde\Sigma},\check{\tilde G}]^K\equiv \tilde{C}$.
The $\xi$-integration delivers
\begin{equation}
\tilde{c}=-\frac{1}{\tau}\left[\langle K\rangle\tilde{g}^K - \langle K\tilde{g}^K\rangle\right],
\end{equation}
with the kernel $K(\varphi-\varphi')\equiv2\pi N_0\tau W(\varphi-\varphi')$,
and where $\langle...\rangle$ is shorthand for angular average.
One then uses the inverse of Eq.~(\ref{postxifinal}) to calculate the corresponding expression
in the standard (``non-tilde'') language.  
We consider separately the effects of spin-orbit ($\boldsymbol{\mathcal A}$)
and of a Zeeman field ($\Psi$), since they add linearly.

First, spin-orbit.
Starting from
\begin{eqnarray}
c 
&=& 
\tilde{c} - \frac{1}{2}
\left\{ 
\frac{\boldsymbol{\mathcal A}\cdot\hat{\bf p}}{p_F}
-
\frac{\boldsymbol{\mathcal A}\cdot\hat{\boldsymbol\varphi}}{p_F}\partial_{\varphi},
\tilde{c}
\right\}
\nonumber\\
\label{trans1}
&=&-\frac{1}{\tau}\left[
\tilde{g} - \langle K\tilde{g}\rangle
\right]
+ 
\nonumber\\
&&
-
\frac{1}{2\tau}
\left\{ 
\frac{\boldsymbol{\mathcal A}\cdot\hat{\bf p}}{p_F}
-
\frac{\boldsymbol{\mathcal A}\cdot\hat{\boldsymbol\varphi}}{p_F}\partial_{\varphi},
\tilde{g} - \langle K\tilde{g}\rangle
\right\},
\end{eqnarray}
the translation from $\tilde{g}$ to $g$ is done by means of Eq.~(\ref{postxi}).
The calculation is easy but some care is needed, so this is done step by step.
First recall that
\begin{equation}
\tilde{g}-
\frac{1}{2}
\left\{ 
\frac{\boldsymbol{\mathcal A}\cdot\hat{\bf p}}{p_F}
-
\frac{\boldsymbol{\mathcal A}\cdot\hat{\boldsymbol\varphi}}{p_F}\partial_{\varphi},
\tilde{g}
\right\}
=
g,
\end{equation}
so that Eq.~(\ref{trans1}) becomes
\begin{eqnarray}
\label{trans2}
c &=&
-\frac{1}{\tau}\left[g - \langle K\tilde{g}\rangle\right]
+
\nonumber\\
&&
+\frac{1}{2\tau}
\left\{
\frac{\boldsymbol{\mathcal A}\cdot\hat{\bf p}}{p_F}
-
\frac{\boldsymbol{\mathcal A}\cdot\hat{\boldsymbol\varphi}}{p_F}\partial_{\varphi},
\langle Kg\rangle
\right\}.
\end{eqnarray}
Then consider the $\langle K\tilde{g}\rangle$ term, where $K=K(\varphi-\varphi')$  
\begin{eqnarray}
\langle K\tilde{g}\rangle 
&=&
\int\,\frac{{\rm d}\varphi'}{2\pi} g(\varphi')+
\nonumber\\
&&
+
\frac{1}{2}
\int\,\frac{{\rm d}\varphi'}{2\pi}
\left\{
\frac{\boldsymbol{\mathcal A}\cdot\hat{\bf p}'}{p_F}K +
\frac{\boldsymbol{\mathcal A}\cdot\hat{\boldsymbol\varphi}'}{p_F}K,
\partial_{\varphi'}g(\varphi')
\right\}
\nonumber\\
&=&
\langle Kg\rangle +
\nonumber\\
&&
+\frac{1}{2}
\int\,\frac{{\rm d}\varphi'}{2\pi}
\left\{
\left[
\frac{\boldsymbol{\mathcal A}\cdot\hat{\bf p}'}{p_F} 
+ 
\frac{\boldsymbol{\mathcal A}\cdot\left(\partial_{\varphi'}\hat{\boldsymbol\varphi}'\right)}{p_F}
\right]K + 
\right.
\nonumber\\
&&
\left.
+
\frac{\boldsymbol{\mathcal A}\cdot\hat{\boldsymbol\varphi}'}{p_F}\left[\partial_{\varphi'}K\right],
g(\varphi')
\right\}
\nonumber\\
&=&
\langle Kg\rangle
+
\frac{1}{2}
\int\,\frac{{\rm d}\varphi'}{2\pi}
\left\{
\frac{\boldsymbol{\mathcal A}\cdot\hat{\boldsymbol\varphi}'}{p_F},
\left[\partial_{\varphi'}K\right]g(\varphi')
\right\},
\nonumber
\end{eqnarray}
having performed a partial integration and used that 
$g(\varphi=0)=g(\varphi=2\pi),\; \partial_{\varphi'}\hat{\boldsymbol\varphi}'=-\hat{\bf p}'$.
This way Eq.(\ref{trans2}) reads
\begin{eqnarray}
c &=&
-\frac{1}{\tau}\left[
g - \langle Kg\rangle\right] 
+
\frac{1}{2\tau}
\left\{
\frac{\boldsymbol{\mathcal A}\cdot\hat{\bf p}}{p_F},
\langle Kg\rangle
\right\}+ 
\nonumber\\
&&
-
\frac{1}{2\tau}
\int\,\frac{{\rm d}\varphi'}{2\pi}
\left\{
\frac{\boldsymbol{\mathcal A}\cdot\hat{\boldsymbol\varphi}'}{p_F},
\left[\partial_{\varphi'}K\right]g(\varphi')
\right\}+
\nonumber\\
&&
\label{trans3}
-
\frac{1}{2\tau}
\left\{
\frac{\boldsymbol{\mathcal A}\cdot\hat{\boldsymbol\varphi}}{p_F}\partial_{\varphi},
\langle Kg\rangle
\right\}.
\end{eqnarray}
Now work on the last term.  Recall the assumption that the scattering amplitude depend only
on the momentum transfer, i.e. $K({\bf p},{\bf p}')=K({\bf p}-{\bf p}')$.  This implies
\begin{equation}
\label{harmonics}
\frac{\hat{\boldsymbol\varphi}}{p_F}\partial_{\varphi}K
=
-\frac{p_F}{m}\hat{\bf p}\partial_{\xi}K - 
\left[
\frac{p_F}{m}\hat{\bf p}'\partial_{\xi}+
\frac{\hat{\boldsymbol\varphi}'}{p_F}\partial_{\varphi'}
\right]
K.
\end{equation}
From the last term of Eq.~(\ref{trans3}) one therefore has
\begin{eqnarray}
-
\frac{1}{2\tau}
\left\{
\frac{\boldsymbol{\mathcal A}\cdot\hat{\boldsymbol\varphi}}{p_F}\partial_{\varphi},
\langle Kg\rangle
\right\}
=
\frac{1}{2\tau}
\left\{
\frac{\boldsymbol{\mathcal A}\cdot{\bf p}_0}{m},
\langle \partial_{\xi}K g\rangle
\right\}
+
&&
\nonumber\\
\frac{1}{2\tau}
\langle
\left\{
\frac{\boldsymbol{\mathcal A}\cdot{\bf p}_0}{m},
\partial_{\xi}K g
\right\}
\rangle +
&&
\nonumber\\
+
\frac{1}{2\tau}
\int\,\frac{{\rm d}\varphi'}{2\pi}
\left\{
\frac{\boldsymbol{\mathcal A}\cdot\hat{\boldsymbol\varphi}'}{p_F},
\left[\partial_{\varphi'}K\right]g(\varphi')
\right\}.&&
\end{eqnarray}  
Substitution back into Eq.~(\ref{trans3}) gives
\begin{eqnarray}
c &=&
-\frac{1}{\tau}\left[
g - \langle Kg\rangle\right] +
\nonumber\\
&&
+
\frac{1}{2\tau}
\left\{
\frac{\boldsymbol{\mathcal A}\cdot\hat{\bf p}}{p_F},
\langle Kg\rangle
\right\} 
+
\frac{1}{2\tau}
\left\{
\frac{\boldsymbol{\mathcal A}\cdot{\bf p}_0}{m},
\langle \partial_{\xi}K g\rangle
\right\}
\nonumber\\
\label{final}
&&
+
\frac{1}{2\tau}
\langle
\left\{
\frac{\boldsymbol{\mathcal A}\cdot{\bf p}_0}{m},
\partial_{\xi}K g
\right\}
\rangle.
\label{collisionspinorbit}
\end{eqnarray}
This expression can also be obtained by a direct
$\xi$-integration of the collision integral in the standard (``non-tilde'') language.
It agrees with the one appearing in [\onlinecite{shytov2006}] for the case
of parabolic bands when one identifies 
${\bf b}\cdot\boldsymbol{\sigma}/2=\boldsymbol{\mathcal A}\cdot{\bf p}/m$, 
${\bf b}$ being the internal spin-orbit field in the language of Ref.~[\onlinecite{shytov2006}].
Besides the first two terms, in which no spin-orbit contribution appears, 
the third term corresponds
to corrections due to the spin dependent density of states, whereas the fourth and fifth
arise from the energy dependence of the scattering amplitude.
In the notation of [\onlinecite{shytov2006}], the former corrections are due to ${\bf M}^d$
and the latter to ${\bf M}^w$ -- notice that some of the ${\bf M}^d$ and ${\bf M}^w$ 
contributions cancel each other because of Eq.~(\ref{harmonics}).

Let us now consider a Zeeman field described by the scalar potential $\boldsymbol{\Psi}$.
Shifting from covariant to non-covariant
quantities is done according to Eq.~(\ref{postxibis}) and its inverse.
Notice that the kernel $K$ is a function of $\xi$ evaluated at $\xi=\epsilon$.
The energy $\epsilon$ is actually sent to zero during the $\xi-$integration,
though since we now have to shift back to the non-covariant language it is
better to explicitly keep track of this dependency.  
At the end it will be as usual $\xi=\epsilon\rightarrow0$.
From the inverse of Eq.~(\ref{postxibis})
\begin{equation}
\label{noncov1}
c(\epsilon) = \tilde{c}(\epsilon) + \frac{1}{2}\left\{\boldsymbol{\Psi},
\partial_{\epsilon}\tilde{c}(\epsilon)\right\}.
\end{equation}
The first term on the r.h.s. gives
\begin{eqnarray}
\tilde{c} &=& -\frac{1}{\tau}\left[\langle K\rangle g-\langle Kg\rangle\right]+
\nonumber\\
&&
- 
\frac{1}{2\tau}
\left[\langle K\rangle\left\{\boldsymbol{\Psi},g'\right\}-
\langle K\left\{\boldsymbol{\Psi},g'\right\}\rangle
\right],
\end{eqnarray}  
whereas the second one leads to
\begin{equation}
\partial_{\epsilon}\tilde{c} = -\frac{1}{\tau}\left[\langle K'\rangle\tilde{g} +\langle K\rangle\tilde{g}'
-\langle K'\tilde{g}\rangle-\langle K\tilde{g}'\rangle\right]
\end{equation}
where $K'=\partial_{\epsilon}K,\tilde{g}'=\partial_{\epsilon}\tilde{g}$.
Plugging both expressions into Eq.~(\ref{noncov1}) one has
\begin{eqnarray}
c(\epsilon) &=& -\frac{1}{\tau}\left[\langle K\rangle g -\langle Kg\rangle\right]+
\nonumber\\
&&
+\frac{1}{2\tau}
\left[\langle K\rangle\left\{\boldsymbol{\Psi},g'\right\}-
\langle K\left\{\boldsymbol{\Psi},g'\right\}\rangle
\right]+
\nonumber\\
&&
-\frac{1}{2\tau}
\left[
\langle K'\rangle\left\{\boldsymbol{\Psi},g\right\}+
\langle K\rangle\left\{\boldsymbol{\Psi},g'\right\}+
\right.
\nonumber\\
&&
\quad\quad
\left.-
\langle K'\left\{\boldsymbol{\Psi},g\right\}\rangle-
\langle K\left\{\boldsymbol{\Psi},g'\right\}\rangle
\right]
\nonumber\\
&=& -\frac{1}{\tau}\left[\langle K\rangle g -\langle Kg\rangle\right]+
\nonumber\\
&&
-
\frac{1}{2\tau}
\left\{
\boldsymbol{\Psi},\langle K'\rangle g-\langle K'g\rangle
\right\}
\label{collisionzeeman}
\end{eqnarray}
with $g=g(\epsilon), g'=\partial_{\epsilon}g$ and $K,K'$ 
are evaluated at the Fermi surface $\xi=\epsilon=0$.
The full expression for the scattering kernel in the presence of
spin-orbit coupling and a Zeeman field is given by the sum of
Eq.~(\ref{collisionspinorbit}) and Eq.~(\ref{collisionzeeman}).
It leads to results in agreement with Ref.~[\onlinecite{engel2007b}].

\bibliography{boltzmann_biblio}

%
\end{document}